\documentclass[aps, prd, superscriptaddress, preprintnumbers, twocolumn, nofootinbib]{revtex4}

\usepackage{amsmath}	
\usepackage{amsthm}	
\usepackage{amssymb}	
\usepackage{datetime}
\usepackage{graphicx}
\usepackage{color}
\usepackage{verbatim}
\usepackage{multirow}

\usepackage[bookmarks, breaklinks, plainpages=false, pdfpagelabels, colorlinks]{hyperref} 
	\hypersetup{linkcolor=darkblue,citecolor=darkblue,filecolor=darkblue,urlcolor=darkblue} 

\definecolor{grey}{rgb}{0.4,0.4,0.4}
\definecolor{dullmagenta}{rgb}{0.4,0,0.4}
\definecolor{darkblue}{rgb}{0,0,0.4}
\definecolor{midblue}{rgb}{0,0,0.5}
\definecolor{midred}{rgb}{0.5,0,0}
\definecolor{orange}{rgb}{1,0.5,0}
\definecolor{lightbrown}{rgb}{0.75,0.5,0.25}
\definecolor{tan}{cmyk}{0.14,0.42,0.56,0}
\definecolor{djunglegreen}{cmyk}{0.99,0,0.52,0}
\definecolor{lightgreen}{rgb}{0,1,0}
\definecolor{olivegreen}{cmyk}{0.64,0,0.95,0.40}
\definecolor{midgreen}{rgb}{0.0,0.675,0.0}
\definecolor{darkgreen}{rgb}{0,0.5,0}

\newcommand{\q}{\quad}
\newcommand{\qq}{\qquad}

\newcommand{\vs}{\vspace}
\newcommand{\hs}{\hspace}
\renewcommand{\.}{\hspace{0.5mm}}

\newcommand{\ra}{\ensuremath{\rightarrow}}

\newcommand{\grm}{\ensuremath{\mathrm{g}}}

\newcommand{\Ocal}{\ensuremath{\mathcal{O}}}

\newcommand{\mnus}{\ensuremath{m_{\nu_{\!s}}}}
\newcommand{\vrms}{\ensuremath{v_{\rm RMS}}}
\newcommand{\sinsquaretwotheta}{\ensuremath{\sin^{2}( 2\mspace{1mu}\Theta )}}

\newcommand{\kmps}{\ensuremath{\rm km/s}}
\newcommand{\cm}{\ensuremath{\rm cm}}
\newcommand{\AU}{\ensuremath{\rm AU}}

\newcommand{\eV}{\ensuremath{\mathrm{eV}}}
\newcommand{\keV}{\ensuremath{\mathrm{keV}}}
\newcommand{\MeV}{\ensuremath{\mathrm{MeV}}}
\newcommand{\GeV}{\ensuremath{\mathrm{GeV}}}
\newcommand{\TeV}{\ensuremath{\mathrm{TeV}}}

\newcommand{\eg}{e.g.}
\newcommand{\ie}{i.e.}

\newcommand{\cf}{cf.} 

\let\baraccent=\= 
\renewcommand{\=}[1]{\stackrel{#1}{=}} 

\theoremstyle{definition}

\theoremstyle{remark}

\settimeformat{ampmtime}

\usepackage{float}

\begin{document}

\title{Signatures of compact halos of sterile-neutrino dark matter}

\author{Florian K{\"u}hnel}
\email{kuhnel@kth.se}
\affiliation{Department of Physics,
	School of Engineering Sciences,\\
	KTH Royal Institute of Technology,
	AlbaNova University Center,\\
	Roslagstullsbacken 21,
	SE--106\.91 Stockholm,
	Sweden}
\affiliation{The Oskar Klein Centre for Cosmoparticle Physics,
	AlbaNova University Center,\\
	Roslagstullsbacken 21,
	SE--106\.91 Stockholm,
	Sweden}

\author{Tommy Ohlsson}
\email{tohlsson@kth.se}
\affiliation{Department of Physics,
	School of Engineering Sciences,\\
	KTH Royal Institute of Technology,
	AlbaNova University Center,\\
	Roslagstullsbacken 21,
	SE--106\.91 Stockholm,
	Sweden}
\affiliation{The Oskar Klein Centre for Cosmoparticle Physics,
	AlbaNova University Center,\\
	Roslagstullsbacken 21,
	SE--106\.91 Stockholm,
	Sweden}
\affiliation{University of Iceland, 
	Science Institute, 
	Dunhaga 3, 
	IS--107 Reykjavik, 
	Iceland}

\date{\formatdate{\day}{\month}{\year}}

\begin{abstract}
We investigate compact halos of sterile-neutrino dark matter and examine observable signatures with respect to neutrino and photon emission. Primarily, we consider two cases: primordial black-hole halos and ultracompact minihalos. In both cases, we find that there exists a broad range of possible parameter choices such that detection in the near future with x-ray and gamma-ray telescopes might be well possible. In fact, for energies above $10\,\TeV$, the neutrino telescope {\it IceCube} would be a splendid detection machine for such macroscopic dark-matter candidates.\\
\end{abstract}

\maketitle

\section{Introduction}
\label{sec:Introduction}
\vs{-3mm}

In the present standard model of cosmology, approximately $25\,$\% of the energy density of the Universe is constituted in the form of a pressureless, nearly perfect fluid of nonrelativistic objects, so-called (cold) dark matter. Many potential dark-matter candidates have been proposed so far. Perhaps the most studied of those is a hypothetical heavy particle beyond the standard model of particle physics, which only weakly interacts with the other standard-model particles, called a WIMP for short. However, it has been realized that dark matter might also exist, possibly partly, in the form of macroscopic objects. Two of those, primordial black holes (PBHs) \cite{1967SvA....10..602Z, Carr:1974nx} and ultracompact minihalos (UCMHs) \cite{Ricotti:2009bs}, will be discussed in greater detail below.\footnote{There exists yet another possibility for macroscopic dark matter, namely in the form of nuclear-density objects (\cf~Refs.~\cite{Witten:1984rs, Lynn:1989xb}).}

On the particle side, amongst many variants of what the dark matter could be, one quite promising possibility is that of so-called sterile neutrinos (see Ref.~\cite{Dodelson:1993je} for an early discussion on their role as dark-matter components and Refs.~\cite{Boyarsky:2009ix, Adhikari:2016bei, Abazajian:2017tcc, Bernal:2017kxu} for more recent reviews on their role in cosmology and astrophysics in general). Sterile-neutrino dark matter, which will be discussed in more detail in the next section, really stands for a whole class of dark-matter models, ranging from one to an, in principle, arbitrary number of additional neutrinos with a broad range of masses from a few $\eV$ to even far above the $\TeV$ scale (\cf~Ref.~\cite{Adhikari:2016bei}) and so-called mixing, determining to which degree the active and sterile neutrino flavor eigenstates constitute the neutrino mass eigenstates.\footnote{Increasing the number of sterile neutrinos, the number of model parameters raises significantly, and hence, the constraints on those become weaker. In view of the possibility to serve as a solution to possible small-scale issues of the standard model of cosmology (such as the so-called missing-satellite \cite{Bode:2000gq} or the too-big-to-fail problem \cite{Lovell:2011rd}), emphasis has mostly been put on investigating cases with up to three sterile neutrinos, where one of them has a mass in the $\keV$ range.}

Combining the two scenarios mentioned above, \ie~particle and macroscopic dark matter, to one consisting of sterile neutrinos aggregated to compact massive structures{---}either in the form of dark-matter halos around PBHs or in the form of UCMHs{---}is very tempting. On the one hand, a small fraction of PBHs could provide excellent seeds for the super-massive black holes in the centers of galaxies \cite{Bean:2002kx}. On the other hand, in view of the fact that it appears difficult, although not impossible, to have the entire dark matter in the form of PBHs or UCMHs (\cf~Ref.~\cite{Carr:2016drx} for a recent article including a summary of relevant constraints), sterile neutrinos appear to date as vital supplementary and major candidates for the dark matter. The derivation of observable signatures for such a combined scenario is the aim of this paper.
\vs{-3mm}

\section{Sterile Neutrinos}
\label{sec:Sterile-Neutrinos}
\vs{-3mm}

As mentioned above, these are hypothetical types of neutrinos that do not interact via any of the fundamental forces in Nature, except for the gravitational force, which makes them even harder to detect than ordinary neutrinos. Despite the fact that sterile neutrinos do not interact via the weak force, since they are inherently right-handed, they can still mix with active neutrinos, which also means that oscillations between active and sterile neutrinos can occur. The mixing between active and sterile neutrinos is quantified by the active-sterile neutrino mixing $\sinsquaretwotheta$, $\Theta$ being the corresponding mixing angle. Furthermore, the number of sterile neutrinos does not affect the measurements of the lifetime of the Z boson, since sterile neutrinos do not interact via the weak force.

The mass of sterile neutrinos is generated by a Majorana mass term and is hence not bounded by the electroweak energy scale. Also, there are no model-independent bounds on the mass of sterile neutrinos, which is denoted by $\mnus$. Therefore, in this work, since their production mechanism is basically unknown, we will take a phenomenological approach and do not restrict ourselves to sterile neutrinos of any particular mass (but we mostly use a mass of $\mnus = 10\,\keV$ as a reference value). This will allow us to probe a wide class of scenarios and even open up the detection possibilities with neutrino telescopes, which are typically not sensitive to decays of light (\eg~\eV-massed) sterile neutrinos.

Most observational investigations of sterile neutrinos as a dark-matter candidate utilize the study of keV x-rays from decays of sterile neutrinos to active neutrinos. These have been performed with a number of observations, such as \emph{Suzaku} \cite{Tamura:2014mta} or \emph{NuSTAR} \cite{Perez:2016tcq} (see also Refs.~\cite{Adhikari:2016bei, Abazajian:2017tcc} and references therein).

\section{Primordial Black Holes}
\label{sec:Primordial-Black-Holes}
\vs{-3mm}

These are black holes which have been produced in the very early Universe. Ever since PBHs were first postulated, they have received considerable attention \cite{1967SvA....10..602Z, Carr:1974nx}. The interest in them constituting (parts of) the dark matter \cite{1975Natur.253..251C} has been revived recently \cite{Carr:2009jm, Bird:2016dcv, Carr:2016drx, Clesse:2016vqa, Green:2016xgy, Kuhnel:2017pwq, Carr:2017jsz, Carr:2017edp}, in particular through the gravitational-wave discovery of black-hole binary mergers \cite{Abbott:2016blz, Abbott:2016nmj}. The possible PBH formation mechanisms are very diverse and there is a plethora of scenarios, which lead to their formation. All of these have in common that they require a mechanism to generate large overdensities, specified by the density contrast $\delta \equiv \delta \rho / \rho$ with $\rho$ denoting the background energy density and $\delta \rho$ the local overdensity.

Often these overdensities are of inflationary origin \cite{Hodges:1990bf, Carr:1993aq, Ivanov:1994pa}. When reentering the cosmological horizon, they collapse if they are larger than a given threshold $\delta_{c}$, which is medium-dependent. For the most relevant case of radiation domination, it is $\delta_{c} \approx 0.45$ (\cf~Ref.~\cite{Musco:2012au}). Here, the case of radiation domination is the one most often considered in the literature. Other scenarios for PBH formation exist, such as those where the sources of the inhomogeneities are first-order phase transitions \cite{Jedamzik:1996mr}, bubble collisions \cite{Crawford:1982yz, Hawking:1982ga}, collapse of cosmic strings \cite{Hogan:1984zb, Hawking:1987bn}, necklaces \cite{Matsuda:2005ez}, or domain walls \cite{Berezin:1982ur}.

With the vast amount of mentioned formation mechanisms, PBHs might have been produced in any possible abundance with masses ranging from about the Planck mass up to twenty orders of magnitude above the solar mass. To date, it is entirely unclear which, if any, of these mechanisms that have been active. In this work, we will therefore take a pragmatic approach and be ignorant to the precise origin of PBHs and pose constraints on a given abundance at a given mass.

Furthermore, we will consider a scenario of two-component dark matter with a small fraction of PBHs and a large complementary fraction of sterile neutrinos forming halos around them. Typically, the mass $M$ contained in the halo is much larger than that of the contained black hole itself $M_{\rm BH}$, \ie~$M \gg M_{\rm BH}$ (see Ref.~\cite{Ricotti:2007au}). We will just refer to these objects as compact halos.

In the following, we will assume that these compact halos constitute a fraction $f_{\rm DM} = 1$ of the dark matter.\footnote{Generalizing to $f_{\rm DM} \leq 1$, a constraint on $f_{\rm DM}$ can easily be set using nonobservation of sterile-neutrino decay signatures.} Since we are dealing with decays, and moreover, for all practical detection purposes, these compact halos are pointlike, and hence, their halo profiles do not matter; the decay characteristics will be {\it entirely} determined by the sterile-neutrino mass $\mnus$, the sterile-active neutrino mixing angle $\Theta$, and the mass of the compact halo $M$. Below, we will elaborate more on the concrete radiation mechanism.

\section{Ultracompact Minihalos}
\label{sec:Ultra--Compact-Mini--Halos}
\vs{-3mm}

Ricotti {\it et al.}~\cite{Ricotti:2009bs} have proposed that large overdensities $3 \cdot 10^{-4} \lesssim \delta_{c} \lesssim 0.3${\,---\,}being therefore below the PBH formation threshold{\,---\,}might form ultradense self-bound structures. In fact, it has been conjectured that, unless PBHs constitute all of the dark matter, UCMHs should be more abundant than PBHs. Recently, various works have investigated constraints on a class of UCMH scenarios \cite{Bringmann:2011ut, Kohri:2014lza, Berezinsky:2014wya, Ali-Haimoud:2015bfg}. Also, their formation mechanisms as well as their halo structure have been thoroughly investigated (\cf~Ref.~\cite{Berezinsky:2013fxa}). Like above for the case of PBHs and for the same reasons, we will neither care about their precise formation mechanism nor about the detailed halo structure, since these will be practically irrelevant for our consideration. So, below we will assume a certain abundance of compact halo objects (which will be either PBHs with halos or UCMHs) of a given $M$.

\section{Radiation from the Compact Halos}
\label{sec:Radiation-from-the-Compact-Halos}
\vs{-3mm}

As regards the emission from the compact halos, there are basically two relevant emission channels: into (a) standard-model neutrinos and (b) photons. Concerning the former, the individual decay rate is given by (\cf~Eq.~(7) in Ref.~\cite{Abazajian:2001vt})
\begin{subequations}
\begin{align}
	\Gamma_{\nu^{}_{\!s} \ra 3\nu}
		&\approx
					8.7 \times 10^{-31}
					\left(
						\frac{ \sinsquaretwotheta }{ 10^{-10} }
					\right)\!
					\left(
						\frac{ \mnus }{ 10\,\keV }
					\right)^{\!5}\,
					{\rm s}^{-1}
					\, ,
					\label{eq:Gamma-nu}
\end{align}
while the decay rate of the latter is about a factor of $1/128$ smaller (\cf~Eq.~(10) in Ref.~\cite{Abazajian:2001vt})
\begin{align}
	\Gamma_{\nu^{}_{\!s} \ra \nu\gamma}
		&\simeq
					\frac{ 1 }{ 128 }\,
					\Gamma_{\nu^{}_{\!s} \ra 3\nu}
					\, .
					\label{eq:Gamma-gamma}
\end{align}
\end{subequations}
Nevertheless, for most cases, the photon emission is by far the most relevant one, as for instance the neutrino telescope {\it IceCube} can only observe neutrinos above an energy of $\Ocal(100)\.\GeV$ \cite{Ahrens:2003ix}.

In order to obtain the total decay rates $\Gamma^{\rm total}_{\nu^{}_{\!s} \ra \nu\gamma}$ and $\Gamma^{\rm total}_{\nu^{}_{\!s} \ra 3\nu}$ for a given compact halo of mass $M$, the individual decay rates in Eqs.~\eqref{eq:Gamma-nu} and \eqref{eq:Gamma-gamma} have to multiplied by the number $N$ of sterile neutrinos within this halo, \ie~$N \simeq M / \mnus$, since $M \gg M_{\rm BH}$. Furthermore, given a homogeneous local dark-matter density of $\rho_{\rm DM} \approx 0.3\,\GeV / {\rm cm}^{3}$, $M$ also determines the average distance between two halos
\begin{align}
	d
		&\approx
					1.2 \times 10^{8}
					\left( \frac{ M }{ 1\.\grm } \right)^{\frac{1}{3}}
					\, {\cm}
		\approx
					8.2 \times 10^{-6}
					\left( \frac{ M }{ 1\.\grm } \right)^{\frac{1}{3}}
					\, \AU
					\, .
					\label{eq:d-average}
\end{align}
In Table~\ref{tab:M-N-d-Gamma-g-Gamma-nu}, the characteristics mentioned above are shown for a few selected values of $M$, for fixed $\mnus$ and $\Theta$ (see table caption).

\begin{table}
	\begin{center}
		\caption{Values for the mass (in g) of the compact halos with associated values for 
				the number of sterile neutrinos they contain, their average distance (in AU) as
				well as the rates of photon ($\gamma$) 
				and neutrino ($\nu$) emission. 
				Here, we assume $\mnus = 10\,\keV$ and $\sinsquaretwotheta = 10^{-11}$. 
				Note that the number of sterile neutrinos is inversely proportional to $\mnus$; 
				the rates depend on this quantity to the fifth power, \cf~Eqs.~(\ref{eq:Gamma-nu}) 
				and (\ref{eq:Gamma-gamma}), 
				and hence can be increased tremendously for larger values of $\mnus$.
				\vs{-2.5mm}}
		\begin{tabular}{ l c c c c}
			\hline
			\hline
			\;$M$\./\.g
			& \qq\;\,$N$
			& \q\;\;\.$d$\./\.AU
			& \q\;$\Gamma^{\rm total}_{\nu^{}_{\!s} \ra \nu\gamma}$\./\.s
			& \q\;$\Gamma^{\rm total}_{\nu^{}_{\!s} \ra 3\nu}$\./\.s\q\\
			\hline
			$\hs{1.5mm}10^{-3}$				& \q\;\;$6 \times 10^{25}$					& \q\;\;$8 \times 10^{-7}$	
			& \q\;\;$4 \times 10^{-6}$			& \q\;\;$5 \times 10^{-4}$\\[0.5mm]
			$\hs{1.5mm}1$					& \q\;\;$6 \times 10^{28}$					& \q\;\;$8 \times 10^{-6}$	
			& \q\;\;$4$						& \q\;\;$500$\\[0.5mm]
			$\hs{1.5mm}10^{3}$				& \q\;\;$6 \times 10^{31}$					& \q\;\;$8 \times 10^{-5}$	
			& \q\;\;$4 \times 10^{6}$			& \q\;\;$5 \times 10^{8}$\\[0.5mm]
			$\hs{1.5mm}10^{6}$				& \q\;\;$6 \times 10^{34}$					& \q\;\;$8 \times 10^{-4}$	
			& \q\;\;$4 \times 10^{12}$			& \q\;\;$5 \times 10^{14}$\\[0.5mm]
			$\hs{1.5mm}10^{9}$				& \q\;\;$6 \times 10^{37}$					& \q\;\;$8 \times 10^{-3}$	
			& \q\;\;$4 \times 10^{18}$			& \q\;\;$5 \times 10^{20}$\\[0.5mm]
			$\hs{1.5mm}10^{12}$			& \q\;\;$6 \times 10^{40}$					& \q\;\;$0.08$	
			& \q\;\;$4 \times 10^{24}$			& \q\;\;$5 \times 10^{26}$\\[0.5mm]
			$\hs{1.5mm}10^{15}$			& \q\;\;$6 \times 10^{43}$					& \q\;\;$0.8$	
			& \q\;\;$4 \times 10^{30}$			& \q\;\;$5 \times 10^{32}$\\[0.5mm]
			$\hs{1.5mm}10^{18}$			& \q\;\;$6 \times 10^{46}$					& \q\;\;$8$	
			& \q\;\;$4 \times 10^{36}$			& \q\;\;$5 \times 10^{38}$\\[0.5mm]
			$\hs{1.5mm}10^{21}$			& \q\;\;$6 \times 10^{49}$					& \q\;\;$80$	
			& \q\;\;$4 \times 10^{42}$			& \q\;\;$5 \times 10^{44}$\\[0.5mm]
			$\hs{1.5mm}10^{24}$			& \q\;\;$6 \times 10^{52}$					& \q\;\;$800$	
			& \q\;\;$4 \times 10^{48}$			& \q\;\;$5 \times 10^{50}$\\[0.5mm]
			\hline
			\hline
		\end{tabular}
		\label{tab:M-N-d-Gamma-g-Gamma-nu}
	\end{center}
\end{table}

Now, the idea is that with a given probability a compact halo passes by within a certain distance to the detector and generates a much larger flux through its effective area than could come from any background or from any of the more remote sources, like decaying sterile neutrinos in a distant galaxy.\footnote{For all considered parameter ranges, the amplification from a compact halo object flying near by (and much nearer than their average distance) the detector can easily exceed any flux from other compact halo objects both in the background and from a distant localized source. To be definite, consider the extreme case of a compact halo object being directly at the detector (and hence $\Ocal( 1 )$ of its radiation is going through the detector area $A_{\rm eff}$), as well as be extremely conservative and take $A_{\rm eff}$ to be equal to $10^{8}\.{\rm cm}^{2}$. Then, for the case of compact halo objects of $M = 10^{15}\.{\rm g}$ with average distance of $d = 0.8\.{\rm AU}$ (\cf~Table~\ref{tab:M-N-d-Gamma-g-Gamma-nu}), the amplification with regard to the flux generated by this object is about $( \pi d^{2} ) / A_{\rm eff} \approx 10^{20}$ times larger than what such an object would generate at $d = 0.8\.{\rm AU}$. Taking the accumulated radiation of all compact halo objects into account brings this ratio down, but nowhere near 1. The same holds even more true for all any distant localized source, due to the $r^{-2}$ decline of the flux.} For definiteness, let us assume a locally homogeneous spatial distribution of compact halos and a Maxwellian velocity distribution:
\begin{align}
	g( v )
		&\equiv
					\left(
						\frac{ 3 }{ 2 \pi\.\vrms^{2} }
					\right)^{\frac{3}{2}}
					\exp\!
					\left(
						-
						\frac{ 3\.v^{2} }{ 2\.\vrms^{2} }
					\right)
					\label{eq:Maxwellian-velocity-distribution}
\end{align}
with the root-mean square velocity $\vrms$.

By virtue of the total decay rates mentioned above, for a given value of $M$, there exists a given distance $r_{\Phi}$ from the detector such that a certain flux $\Phi^{}_{\!A}$ through the effective detector area $A$ is observed, namely
\begin{align}
	r_{\Phi}
		&\simeq
					\sqrt{	
						\frac{ A }{ 4\.\pi\.\Phi^{}_{\!A} }
					}
					\cdot
					\begin{cases}
						\sqrt{N\.\Gamma_{\nu^{}_{\!s} \ra \nu\gamma}\,}\\[1mm]
						\sqrt{N\.\Gamma_{\nu^{}_{\!s} \ra 3\nu}\,}
					\end{cases}
					\notag
					\\[2mm]
		&\approx
					\left(
						\frac{ \Phi^{}_{\!A} }{ 1\.A^{-1}\.{\rm s}^{-1} }
					\right)^{-\frac{1}{2}}
					\cdot
					\begin{cases}
						1.7 \times 10^{3}\\[1mm]
						2.0 \times 10^{4}
					\end{cases}
					\notag
					\\[0.5mm]
		&\phantom{\approx\;}
					\times
					\sqrt{
					\left(
						\frac{ A }{ 1\,\cm^{2} }
					\right)\!
					\left(
						\frac{ M }{ 1\,\grm }
					\right)\!
					\left(
						\frac{ \mnus }{ 1\,\keV }
					\right)^{\!4}\.
					\sinsquaretwotheta
					}\;
					\cm
					\, ,
					\label{eq:rPhi}
\end{align}
where, as above, $N \simeq M / \mnus$ is the number of sterile neutrinos within each compact halo.

\begin{figure}
	\centering
	\includegraphics[width=0.375\textwidth]{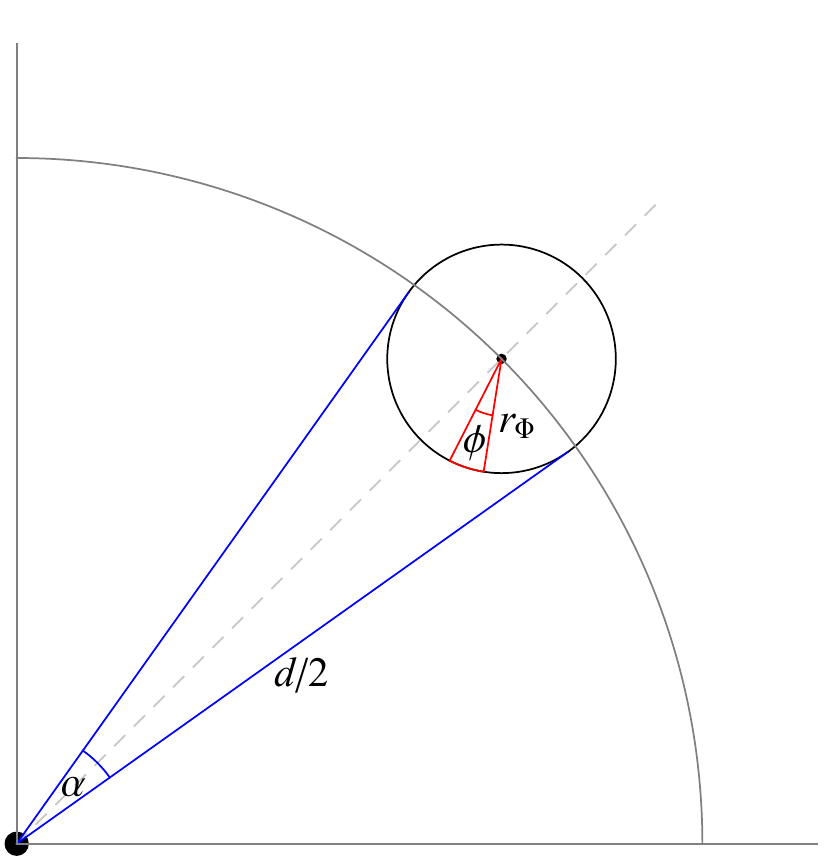}
	\caption{
		Two-dimensional illustration for a geometric derivation of the probability $P$,
		\cf~Eq.~\eqref{eq:P}. 
		A situation of a compact halo (lower-left corner), which moves into a region containing
		the detector, is shown.
		For a given flux $\Phi$, its sphere (radius $r_{\Phi}$, at $d / 2$ away from the halo) 
		subtends an angle $\alpha \equiv \arcsin( 2\.r_{\Phi} / d )$ as seen from the compact halo.
		Also, a fraction of the sphere accessible by the detector through one of its opening angles
		(here:~$\phi$) is displayed.
		}
	\label{fig:Geometry}
\end{figure}

Given the average distance $d$ and velocity $v$ of these halos, one can easily (in fact, purely geometrically\footnote{Since we are concerned with our very local neighborhood, we assume homogeneously-distributed dark matter (\ie~compact halos), one may imagine them as being placed on a uniform grid (of edge size equal to $d$) with Earth somewhere within this grid, say, in the middle of one of the grid cells for definiteness. At the corners of each of these cells, there are all together eight halos, which move in random directions. Now, each halo has eight octants to move to, so on average one of them will move into the octant containing Earth. Therefore, let us focus on this one. The probability for such a compact halo to intersect one of the flux spheres mentioned above is then just given by the ratio of the angles this sphere subtends (from the halo's perspective), divided by $\pi^{2} / 4$ (from the octant) {\it times} the fraction of the sphere intersected by the cone given by the detector opening angles.}, see Fig.~\ref{fig:Geometry}) compute the probability $P$ that a certain such flux sphere is intersected, \ie
\begin{align}
	P
		&=
					\frac{ \theta \cdot \phi }{ \pi \cdot 2\.\pi }\,
					\frac{ \alpha^{2} }{ \pi^{2} / 4 }					
		\simeq
					\frac{ 16 }{ \pi^{2} }
					\left(
						\frac{ \theta\,\phi }{ 2\.\pi^{2} }
					\right)
					\arcsin^{2}\mspace{-3mu}
					\left(
						\frac{ 2\.r_{\Phi} }{ d }
					\right)
					\, .
					\label{eq:P}
\end{align}
Above, $\theta \in [0,\.\pi]$ and $\phi \in [0,\.2\pi)$ are the opening angles of a detector. The probability $P$ in Eq.~\eqref{eq:P} can then be translated to the time it takes to observe a given flux. A$\mspace{-1mu}$veraged over the velocity distribution in Eq.~\eqref{eq:Maxwellian-velocity-distribution}, it reads
\begin{align}
	T_{f}
		&\simeq
					\left(
						\frac{ 1 }{ P }
						-
						\frac{ 1 }{ 2 }
					\right)
					\frac{
						d
						-
						2\.r_{\Phi}
					}
					{
						\vrms
					}\;
					\notag
					\\[2mm]
		&\approx
					2.5 \times 10^{-8}
					\left(
					\frac{
						d
						-
						2\.r_{\Phi}
					}
					{
						1\,{\rm km}
					}
					\right)\!
					\left(
						\frac{ \vrms }{
							200\,{\rm km / s}
						}
					\right)^{\!\!-1}
					\notag
					\\[1mm]
		&\phantom{\approx\;}
					\times
					\left[
						\frac{ \pi^{2} }{ 8 }\!
						\left(
							\!\frac{ \theta\,\phi }{ 2\.\pi^{2} }\!
						\right)^{\!\!-1}
						\arcsin^{-2}\mspace{-3mu}
						\left(
							\frac{ 2\.r_{\Phi} }{ d }
						\right)
						-
						1
					\right]
					\, {\rm s}
					\, ,
					\label{eq:Tf}
\end{align}
which is valid as long as $d$ is larger than $2\.r_{\Phi}$; if it is smaller, the sought-for flux $\Phi$ would be below background level.

\begin{figure}
	\centering
	\includegraphics[width=0.5\textwidth]{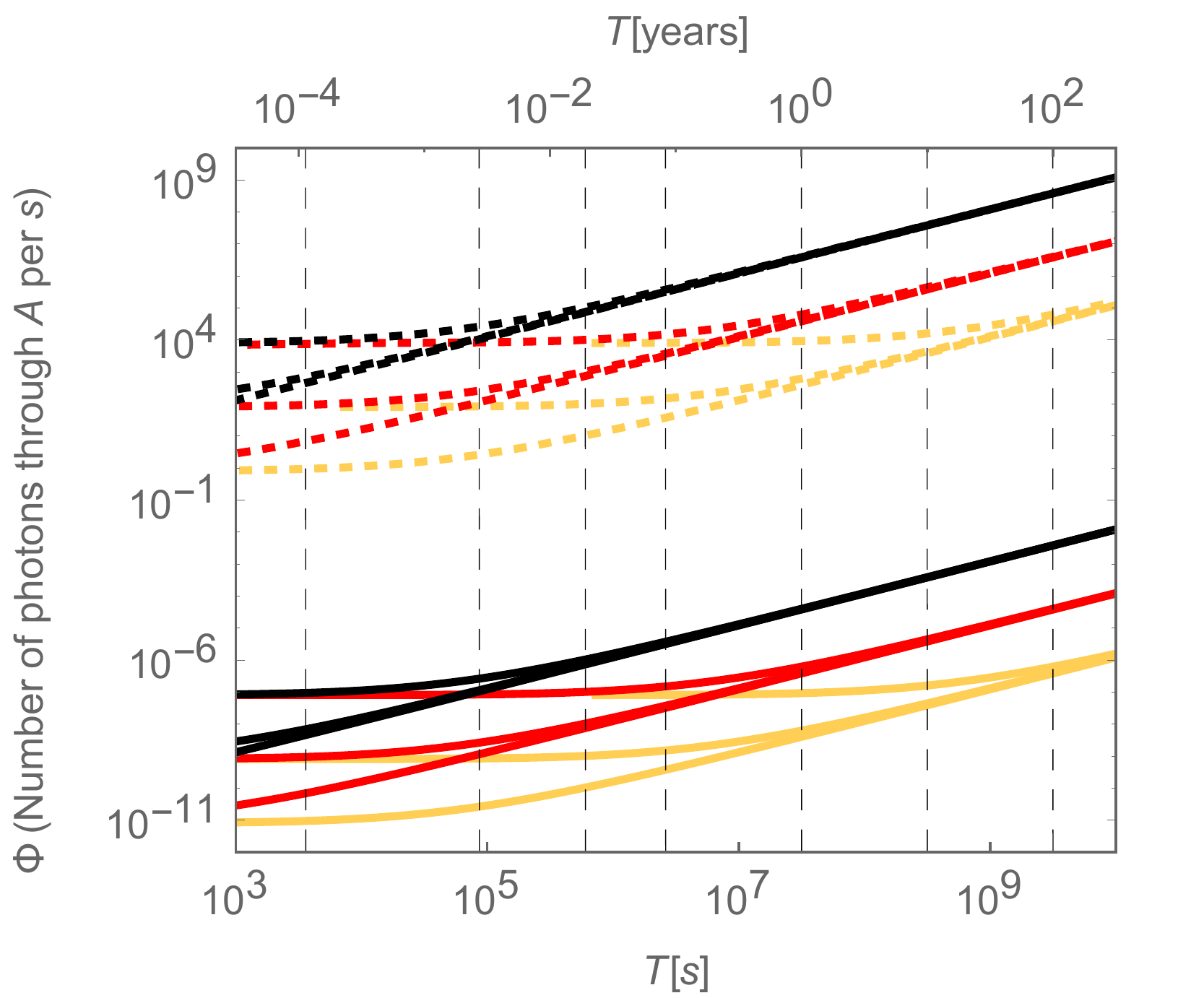}
	\caption{
		Photon flux $\Phi^{}_{\!A}$ through the effective detector area $A$ as a function of 
		the observation time $T$ necessary to achieve this flux.
		The different colors depict different root-mean square velocities $\vrms$
		assuming (for definiteness) a Maxwellian distribution of compact-halo velocities. 
		We visualize results for a hypothetical detector with an
		effective detection area $A = 10^{5}\,\cm^{2}$ 
		and opening angles of $16^{\circ}$ each.
		The used root-mean square velocities are taken to be 
		$\vrms \in \{ 100,\.200,\.300 \} \,\kmps$, 
		corresponding to yellow, red, and black color coding, respectively.
		Within each of these, results for various compact-halo masses are shown in the range
		$M \in \{ 1,\.10^{6},\.10^{12} \}\,\grm$ (from top to bottom).
		The vertical dashed lines indicate 
		one hour, one day, one week, one month, one year, ten years and 
		hundred years of time (from left to right). 
		The solid curves are results for a sterile-neutrino mass of $\mnus = 10\,\keV$ and 
		a sterile-active neutrino mixing angle $\Theta$ with $\sinsquaretwotheta = 10^{-11}$,
		whereas the dashed curves utilize $\mnus = 100\,\MeV$ and 
		$\sinsquaretwotheta = 10^{-16}$.
		}
	\label{fig:Flux}
\end{figure}

In Fig.~\ref{fig:Flux}, the result of the numerical inversion of Eq.~\eqref{eq:Tf} is presented. The curves in this plot essentially consist of two parts: (i) A constant part to the left of these plots, which corresponds to a background of the integrated photon/neutrino emission surrounding the detector. Here, we use the local dark matter density of $\rho_{\rm DM} \approx 0.3\,\GeV / {\rm cm}^{3}$, and assume that the detector is pointing away from the Galactic center. In the background region, $d$ is smaller than $r_{\Phi}$. The point at which these lengths equal each other is precisely the point when $T_{f}$ in Eq.~\eqref{eq:Tf}, which has been inverted to yield the results in Fig.~\ref{fig:Flux}, vanishes. (ii) A part which increases with mass. This corresponds to the fact that the path of a given compact halo passes through the flux sphere given by $r_{\Phi}$. Furthermore, in Fig.~\ref{fig:Flux}, the various colors denote different root-mean square velocities $\vrms$ (\cf~\cite{Herzog-Arbeitman:2017zbm}), for each of which the results for several values of $M$ are shown (see the caption of Fig.~\ref{fig:Flux} for details). We note that given the already lager velocity spread used, all respective curves only show a minor variation. Also, regardless of the value of $M$, we observe that each curve (for a given value of $\vrms$) converges to a common branch at longer observational times.

Even though the most studied scenarios of sterile-neutrino dark matter focus on mass ranges around a few keV, there is, {\it a priori}, no fundamental reason to reject larger masses{---}even more so as the sterile-neutrino production mechanism is entirely unknown. Furthermore, from Eqs.~\eqref{eq:Gamma-nu} and \eqref{eq:Gamma-gamma}, we see that the decay rates, and hence, the observable signatures, while scaling linearly with $\sinsquaretwotheta$, depend on $\mnus$ to the fifth power. This strong dependence is well visible by comparing the solid curves to the respective dashed ones in Fig.~\ref{fig:Flux}, where the former is for $\mnus = 10\,\keV$ and $\sinsquaretwotheta = 10^{-11}$ and the latter utilizes $\mnus = 100\,\MeV$ and $\sinsquaretwotheta = 10^{-16}$. This results in a shift upwards by a factor of $10^{15}$ from the former to the latter set of curves. It should be noted that the smaller $M$ is, the lower lies the constant (background) part and the earlier it joins the increasing branch. The mentioned shifts of the graphs are the only changes due modifications of the model parameters; the entire functional form stays the same otherwise.

\section{Observability}
\label{sec:Observability}
\vs{-3mm}

The questions of if and how the emission from those compact halos could be detected, depend on both the type of emission and its energy range. Furthermore, besides the photon or neutrino background due to the spatial distribution of compact objects like PBHs with their halos or UCMHs, there are various other astrophysical sources, which constitute a ground level of respective emission.

The relevant photon detectors basically divide into those observing x-rays (such as {\it Suzaku} \cite{Tamura:2014mta}, {\it NuSTAR} \cite{Perez:2016tcq}, and {\it PoGO+} \cite{2017NIMPA.859..125C, Chauvin:2017qnn}) and those measuring gamma-rays (like {\it ACT} \cite{Boggs:2006mh}, {\it AdEPT} \cite{Hunter:2013wla}, and {\it Fermi} \cite{2012ApJS..203....4A}), while the neutrino telescope {\it IceCube} \cite{Aartsen:2016oji} operates from about $100\,\GeV$ and upwards. In Table~\ref{tab:constraint-prospects}, we show results for the values of the threshold compact-halo mass $M_{\rm th}$ for the above-mentioned telescopes. We present our results for an observational time of one year, $\vrms = 200\,{\rm km / s}$, energies $E$ ranging from $1\,\keV$ to $E = 10^{4}\,\TeV$, and five different values for the total decay rate $\Gamma^{\rm total} \in \{ 0.01,\.0.1,\.1,\.10,\.100 \}\cdot\tilde{\Gamma}^{\rm total}$, where the reference decay rate $\tilde{\Gamma}^{\rm total}$ assumes $\mnus = 10\,\keV$ and $\sinsquaretwotheta = 10^{-11}$. For each energy $E$, a given decay rate corresponds to a specific choice of the active-sterile neutrino mixing $\sinsquaretwotheta$, which scales as $\sim E^{-4}$ [\cf~Eq.~\eqref{eq:Gamma-nu}].

Hence, Table~\ref{tab:constraint-prospects} provides constraint prospects for the dark-matter scenario constituted by compact halos of sterile neutrinos such that these objects should be detectable if their masses exceed the values of $M_{\rm th}$ specified in this table. Below these values, the flux received by a detector is smaller than its sensitivity for the assumed values of $\vrms$ and $\sinsquaretwotheta$ and the chosen observational time of one year. A longer observational time increases the probability of detecting a flux peak through increased proximity of the compact halos to the telescope~(\cf~Fig.~\ref{fig:Flux}).

\begin{table}
	\begin{center}
		\caption{Order-of-magnitude detection prospects 
				for the dark-matter scenario constituted by compact halos of sterile neutrinos.
				Results for various telescopes,
				x-ray (top group),
				gamma-ray (middle group), and
				neutrinos (bottom group)
				are presented for an observational time of one year 
				(except for {\it PoGO+} for which we choose two weeks due to technical limitations of this mission)
				and a root-mean square 
				velocity $\vrms = 200\,{\rm km / s}$.
				Energies from $E = 1\,\keV$ to $E = 10^{3}\,\TeV$ have been used.
				For each of these energies, five different decay rates
				$\Gamma^{\rm total} \in \{ 0.01,\.0.1,\.1,\.10,\.100 \}\cdot\tilde{\Gamma}^{\rm total}$
				with the reference $\tilde{\Gamma}^{\rm total}$, assuming
				$\mnus = 10\,\keV$ and $\sinsquaretwotheta = 10^{-11}$,
				have been set and the threshold compact-halo masses $M_{\rm th}$ have been
				determined. 
				Perfect exclusion, \ie~if $M_{\rm th}$ would be smaller than the Planck mass, is
				denoted by a long dash (---).
				\vs{0.5mm}}
		\begin{tabular}{ l c c c c c c}
			\hline
			\hline
			\multicolumn{2}{c}{\underline{\hs{4mm}Decay Rate$_{_{\color{white}0}}$\hs{4mm}}}
			& \multicolumn{5}{c}{$\underline{\hs{14mm}\Gamma^{\rm total} / \tilde{\Gamma}^{\rm total^{^{^{^{}}}}}\hs{14mm}}$}\\
			Telescope
			& 
			& $\;\; 0.01$
			& $\;\; 0.1$
			& $\q\. 1$
			& $\;\; 10$
			& $\;\; 100\;$\\
			\hline
			
			& $E/\eV$
			& \multicolumn{5}{c}{\.Threshold masses $M_{\rm th} / \grm$}\\
			\\[-3mm]
			\;{\it Suzaku}
								&\; $10^{3}$		&\; $10^{36}$		&\; $10^{30}$
								&\; $10^{24}$		&\; $10^{18}$		&\; $10^{12}$\\
			\;{\it NuSTAR}
								&\; $10^{4}$		&\; $10^{33}$		&\; $10^{27}$
								&\; $10^{21}$		&\; $10^{15}$		&\; $10^{9}$\\
			\;{\it PoGO+}
								&\; $10^{5}$		&\; $10^{39}$		&\; $10^{33}$
								&\; $10^{27}$		&\; $10^{21}$		&\; $10^{15}$\\
			\\[-3mm]
			\;{\it ACT}
								&\; $10^{6}$		&\; $10^{30}$		&\; $10^{24}$
								&\; $10^{18}$		&\; $10^{12}$		&\; $10^{6}$\\
			\;\multirow{2}{*}{\vs{-1mm}{\it AdEPT}}	
								&\; $10^{7}$		&\; $10^{33}$		&\; $10^{27}$
								&\; $10^{21}$		&\; $10^{15}$		&\; $10^{9}$\\
								&\; $10^{8}$		&\; $10^{29}$		&\; $10^{23}$
								&\; $10^{17}$		&\; $10^{11}$		&\; $10^{5}$\\
			\;\multirow{4}{*}{\vs{-2.2mm}{\it Fermi}}
								&\; $10^{9}$		&\; $10^{26}$		&\; $10^{20}$
								&\; $10^{14}$		&\; $10^{8}$		&\; $100$\\
								&\; $10^{10}$		&\; $10^{22}$		&\; $10^{16}$
								&\; $10^{10}$		&\; $10^{4}$		&\; $0.01$\\
								&\; $10^{11}$		&\; $10^{20}$		&\; $10^{14}$
								&\; $10^{8}$		&\; $100$			&\; $10^{-4}$\\
								&\; $10^{12}$		&\; $10^{20}$		&\; $10^{14}$
								&\; $10^{8}$		&\; $100$			&\; $10^{-4}$\\
			\\[-3mm]
			\;\multirow{4}{*}{\vs{-2.5mm}{\it IceCube}}
								&\; $10^{13}$		&\; $10^{10}$		&\; $10^{4}$
								&\; $0.01$			&\; ---			&\; ---\\
								&\; $10^{14}$		&\; $10^{6}$		&\; $1$
								&\; ---			&\; ---			&\; ---\\
								&\; $10^{15}$		&\; $10^{3}$		&\; $10^{-3}$
								&\; ---			&\; ---			&\; ---\\
								&\; $10^{16}$		&\; $1$			&\; ---
								&\; ---			&\; ---			&\; ---\\
		\hline
		\hline
		\end{tabular}
		\label{tab:constraint-prospects}
	\end{center}
\end{table}

\section{Summary and Outlook}
\label{sec:Summary-and-Outlook}
\vs{-3mm}

In this paper, we have investigated a class of dark-matter scenarios which consists of sterile neutrinos confined into either halos around primordial black holes or ultracompact minihalos. Special emphasis has been given to estimate possible decay signatures for both photons and neutrinos.

For various parameters (such as the masses of the compact halos), we have derived the observational times necessary to detect these signatures. We have compared these to the detector sensitivities for different energies, taking into account the specific background of photons or neutrinos, respectively. For better comparability, we have studied active-sterile neutrino mixing leading to the same total decay rate for each energy. We have found that almost all state-of-the-art x-ray and gamma-ray telescopes should well be in a position to set bounds on the investigated form of sterile-neutrino compact-halo dark matter. Amongst the photon telescopes, {\it Fermi} should be exceptionally well suited in search of this form of dark-matter objects, regarding both its sensitivity and the width of its detection range. Future surveys might include {\it CTA} \cite{Consortium:2010bc}, which will extend the search range to energies above $10\,\TeV$. Surprisingly, for these energies, the same holds true also for the neutrino telescope {\it IceCube}, which would be one of the most splendid detection machines for such macroscopic dark-matter candidates. Of course, our results only provide first estimates; a proper quantitative analysis is very tempting and should be performed in the future.

\section*{ACKNOWLEDGMENTS}
\vs{-3mm}

It is a pleasure for us to thank Jon Dumm, Chad Finley, Mark Pearce, and Stephan Zimmer for invaluable information regarding instrumental characteristics. We thank the anonymous referee for helpful comments. F.K.~acknowledges support by the Swedish Research Council (Vetenskapsr{\aa}det) through contract No.~638-2013-8993 and the Oskar Klein Centre for Cosmoparticle Physics and T.O.~acknowledges support by the KTH Royal Institute of Technology for a sabbatical period at the University of Iceland.


\end{document}